\documentclass[a4paper,pre,amsmath,amssymb,nofootinbib]{revtex4}

\usepackage{graphicx}
\usepackage{amsmath}
\usepackage{bm}
\renewcommand{\mathbf}{\bm}
\newcommand{\be}{\begin{equation}}
\newcommand{\ee}{\end{equation}}

\begin{document}

\title{Vector modes generated by primordial density fluctuations}

\author{Teresa Hui-Ching Lu}
\email{teresa.huichinglu@gmail.com}

\author{Kishore Ananda}
\email{kishore.ananda@gmail.com}

\author{Chris Clarkson}
\email{chris.clarkson@uct.ac.za}

\date{\today}

\affiliation{Cosmology and Gravity Group, Department of Mathematics
and Applied Mathematics, University of Cape Town, Rondebosch 7701,
Cape Town, South Africa}

\begin{abstract}

While vector modes are usually ignored in cosmology since they are
not produced during inflation they are inevitably produced from the
interaction of density fluctuations of differing wavelengths. This
effect may be calculated via a second-order perturbative expansion. We
investigate this effect during the radiation era. We discuss the
generation mechanism by investigating two scalar modes interacting,
and we calculate the power of vector modes generated by a power-law
spectrum of density perturbations on all scales.

\end{abstract}

\maketitle


\section{INTRODUCTION}

High-precision data from observations of the Cosmic Microwave
Background (CMB)~\cite{WMAP3} and Large Scale Structure
(LSS)~\cite{Tegmark} provide strong evidence for a nearly
spatially flat universe with a primordial spectrum of adiabatic,
Gaussian and nearly scale-invariant density perturbations. The
standard cosmological model provides a remarkable theoretical
basis for these and other observed features of our universe.
Perturbations generated from inflation give a nearly
scale-invariant spectrum of scalar (density) perturbations, and
tensor (gravitational wave) modes, with amplitudes which are
typically within a few orders-of-magnitude of each other. However,
within this paradigm the amplitude of any corresponding vector
modes is zero since a scalar field can't support vector modes at
linear order~\cite{BTW}; even if they were generated during
inflation, vectors decay rapidly after they leave the Hubble
radius during inflation, whereas scalars and tensors are typically
frozen on super-Hubble scales, and only decay when they re-enter
the horizon. A generic prediction of inflation is therefore no
vector modes.

There is an important caveat to this argument. Vector modes are
generated via the non-linear interaction of scalar (and tensor)
perturbations of differing wavelength, and therefore inflation must
generically predict a spectrum of vector modes, but at second-order
in a perturbative expansion. Indeed, observations of the scalar
spectrum requires this to be so independently of whether inflation
is the correct model of the early universe or not. We shall consider
the generation of vectors from scalars in some detail. The analogous
process of gravitational wave generation by scalar-scalar
interaction has been investigated~\cite{TomK, Mat1, Mat2, Mat3,
Noh1, Carb1, Carb2, Nak1, Nak2, East1, ACW, Osano, East2, Stein,
Mat4, Dufaux}, and the work presented here is closely related to
these studies.

We shall principally expand on work of Mollerach \emph{et
al.}~\cite{Mol1}, who considered the effect of secondary vectors on
the CMB. We shall investigate the generation method of vectors from
two scalar modes (we show that unlike gravitational waves, vectors can't be
generated by a single mode), and we shall calculate the power
spectra of the vector part of the metric in the radiation era. Our aim here is to principally discuss how vectors are generated, and get an overall estimate as to the magnitude and distribution of vectors at the end of the radiation era. How the results here relate to observables in the CMB, and the spectrum of vectors today, is left for future work.

Other people have discussed second-order vector modes before. Most
recently, Mena \emph{et al.}~\cite{Mena} considered second-order
vector modes in a collapsing universe. Matarrese~\emph{et
al.}~\cite{Mat5} discussed the generation of primordial magnetic
fields from density perturbations (although see~\cite{Maart1}).
Various other work has discussed vector mode generation on a more
formal level~\cite{Tomita1, Tomita2, Tomita3, Nak1, Nak2, Nakamura3,
Hwang}.

There are a variety of other mechanisms which predict vector modes,
all of which must happen after inflation, and usually predict a
spectrum of modes on small scales. Such sources include cosmic
strings~\cite{Pogosian}, topological defects~\cite{Durrer},
fine-tuned anisotropies in collisionless neutrinos~\cite{Lewis1} and
the presence of an primordial magnetic field~\cite{Sub1, Sub2, Sub3,
Lewis2}. The generated vector modes are highly non-Gaussian.
Second-order inflationary vector modes considered here are also
non-Gaussian and have a $\chi^2$-distribution.

Vector modes are likely to play a more prominent role in cosmology
in the coming years, through their contributions to the
CMB~\cite{Anile,Tomita1, Tomita2, Tomita3, Bartolo1, Bartolo2,
Bartolo3, Mat6}, which will impact on the B-mode polarization and
could be the dominant contribution when compared with the second
order gravitational waves~\cite{Mol1} (assuming one is able to
subtract the lensing signal). They are also crucial in
magneto-genesis~\cite{Dimopoulos, Davis, Bert1, Bert2, Mat5,
Gopal1, Gopal2, Takahashi1, Takahashi2, Sub1, Sub2, Sub3,
Lewis2,Tsagas} as the magnetic field vector has a dominant vector
part. It is therefore appropriate to now consider in more detail the
spectrum of vector modes which we know must exist by virtue of
second-order effects.

The paper is organised as follows. In Sec. II we consider the
formalism for investigating the generation of vectors from
scalars. Then, In Sec. III we discuss the power spectra of vectors
in the radiation era for power-law scalar modes. We also
investigate the generation mechanism of vectors by considering the
interaction of two distinct scalar modes. Finally, we conclude in
Sec. IV.

\section{Generation of vector modes from density perturbations}

We shall consider perturbations of a flat Robertson-Walker background
up to second order. The metric is decomposed as
\begin{equation}
\bar{g}_{\mu \nu} = g_{\mu \nu} + \delta^{(1)} g_{\mu \nu} +
\delta^{(2)} g_{\mu \nu}, \label{perturbed FLRW}
\end{equation}
where Greek indices run from $0\ldots3$ and Latin indices run from
$1\ldots3$. We are only investigating the second order vector modes
sourced by the first order scalar perturbation; therefore, we have
scalar degrees of freedom at first order, $\delta^{(1)} g_{\mu
\nu}$, while the second order perturbations, $\delta^{(2)} g_{\mu
\nu}$, are pure vector modes, and second-order equations are
therefore meaningful when projected out accordingly. Our metric in terms of
conformal time with a longitudinal gauge chosen, is given as
\begin{equation}
\bar{g}_{00} = -a^2 \left( 1 + 2 \Phi^{(1)} \right),\ \
\bar{g}_{0i} = -\frac{1}{2}a^2   S_{i}^{(2)},\ \ \bar{g}_{ij} =
a^2  \left (1 - 2 \Phi^{(1)} \right ) \gamma_{ij},
\label{perturbed metric}
\end{equation}
where $\Phi^{(1)}$ is the first order Bardeen potential, and
$S_{i}^{(2)}$ describes the gauge-invariant~\cite{Mena} second order
vector modes, so that $\partial^{i} S^{(2)}_{i} = 0$. As there is no
ambiguity in what follows we shall drop the order superscripts and
just write $\Phi$ and $S_i$.

\subsection{Density perturbations at linear order}

We shall consider the generation of vectors from scalars
during the radiation era below, but for now
we shall consider the situation where we have a linear constant
equation of state, $c_s^2 = w$ where $w=p/\rho$ and $c_s$ is the
sound speed. We shall assume that the first order matter
perturbations are adiabatic, i.e., that pressure perturbations obey
$\delta^{(1)} p = c_s^2 \delta^{(1)} \rho$. Then the first order
equation of motion for the Bardeen potential in Fourier space
is~\cite{Kodama}
\begin{equation}
\Phi''(\mathbf{k},\eta) + 3 \mathcal{H} (1 + c_s^2)
\Phi'(\mathbf{k},\eta) + c_s^2 k^2 \Phi(\mathbf{k},\eta) = 0,
\label{equation of motion for Bardeen potential}
\end{equation}
where a prime denotes differentiation with respect to conformal time
$\eta$.

In the radiation era, the scale factor, the conformal Hubble rate
and energy density evolve as $a(\eta) = a_0 \frac{\eta}{\eta_0}$,
$\mathcal{H} = \frac{a'}{a} = \frac{1}{\eta}$ and $\rho \propto
\eta^{-4}$, and the general solution to \eqref{equation of motion
for Bardeen potential}, with $c_s^2 = \frac{1}{3}$ is
\begin{equation}
\Phi_r(\mathbf{k}, \eta) = \frac{A_r(\mathbf{k})}{(k \eta)^3}
\left [ \sin \left ( \frac{k \eta}{\sqrt{3}} \right )- \frac{k
\eta}{\sqrt{3}} \cos \left ( \frac{k \eta}{\sqrt{3}} \right )
\right ] + \frac{B_r(\mathbf{k})}{(k \eta)^3} \left [ \frac{k
\eta}{\sqrt{3}} \sin \left ( \frac{k \eta}{\sqrt{3}} \right ) +
\cos \left ( \frac{k \eta}{\sqrt{3}} \right ) \right ].
\label{complete Bardeen potential in radiation}
\end{equation}
We shall ignore the decaying mode~-- that is terms with a $B_r(\mathbf{k})$
coefficient.

Assuming that the fluctuations are Gaussian, we may introduce
Gaussian random variables, $\hat{E}$, with unit variance and the
property
\begin{equation}
\langle \hat{E}^{\ast}(\mathbf{k}_1) \hat{E}(\mathbf{k}_2) \rangle
= \delta^3(\mathbf{k}_1 - \mathbf{k}_2).
\end{equation}
We can then separate the length and directional dependence of
functions of $\mathbf{k}$ and write
$\Phi(\mathbf{k},\eta)=\Phi({k},\eta)\hat E(\mathbf{k})$ and
$A_r(\mathbf{k})=A_r({k})\hat E(\mathbf{k})$.

The power spectrum for the first order scalar perturbation can be
defined through
\begin{equation}
\langle \Phi^{\ast}(\mathbf{k}_1,\eta) \Phi(\mathbf{k}_2,\eta)
\rangle = \frac{2 \pi^2}{k^3} \delta^3(\mathbf{k}_1 -
\mathbf{k}_2) \mathcal{P}_{\Phi}(k, \eta).
\end{equation}
At early times during the radiation era the power spectrum becomes
\begin{eqnarray}
\mathcal{P}_{\Phi}(k) &\simeq& {A}_r(k)^2 \frac{k^3}{486\pi^2}.
\end{eqnarray}
Relating the Bardeen potential to the comoving curvature
perturbation at early times gives us
\begin{eqnarray}\label{ar}
{A}_r(k)^2 &\approx& \frac{216\pi^2}{k^3} \Delta_{\mathcal{R}}^{2}(k),
\end{eqnarray}
where $\Delta_{\mathcal{R}}^{2}$ is primordial power spectrum for
the curvature perturbation $\mathcal{R}$. Current observations show
$\Delta_{\mathcal{R}}^{2} \approx 2.4 \times 10^{-9}$ at a scale
$k_{CMB}=0.002 \mathrm{Mpc}^{-1}$, and is almost independent of
wavenumber on these scales~\cite{WMAP3}.

\subsection{Second-order vector modes}

The vector perturbations at linear order satisfy an evolution
equation and a momentum constraint equation which can be found by
calculating the $i-j$ and $0-i$ parts of the Einstein Field
Equations (EFE's) respectively~\cite{Mena}. In the case of a perfect
fluid (at first order only) there is no source in the vector
evolution equation and it admits solutions proportional to $1/a^2$.
The momentum constraint equation relates the vector perturbation to
the 3-velocity perturbation, which in the perfect fluid case would
be zero. However, the respective equations at second order differ
significantly. Firstly, as we will see, the evolution equation is
sourced, allowing for the generation of vector modes. Secondly,
the momentum constraint no longer excludes the existence of vector
perturbations, provided we are only considering a perfect fluid up
to first order (see~\cite{Mena}).

We calculate the evolution equation for $S_i$ in the usual manner,
by expanding the EFE's up to second order, keeping terms quadratic
in the first order quantities. We start with the trace reversed
EFE's
\begin{equation}
\bar{R}_{\alpha \beta} = 8 \pi G \left ( \bar{T}_{\alpha \beta} -
\frac{1}{2} \bar{g}_{\alpha \beta} \bar{T} \right ) = 8 \pi G
\bar{\Upsilon}_{\alpha \beta}.
\end{equation}
The second order space-space part of the Ricci tensor can be written
as
\begin{equation}
\delta^{(2)} R_{ij} = \delta^{(2)} R^{\mathcal{V}}_{ij} +
\delta^{(2)} R^{(S,S)}_{ij},
\end{equation}
where we have
\begin{equation}
\delta^{(2)} R^{\mathcal{V}}_{ij} = \frac{1}{2} \partial_{(i}
S'_{j)} + \mathcal{H} \partial_{(i} S_{j)},
\end{equation}
which contains only the second order term; and
\begin{equation}
\begin{split}
\delta^{(2)} R^{(S,S)}_{ij} & = \left [ 8 \left ( \frac{a''}{a} +
\mathcal{H}^2 \right ) \Phi^2 + 16 \mathcal{H} \Phi \Phi' + 2
\Phi'^2 + 2 \Phi \Phi'' + 2 \Phi\ \nabla^2 \Phi + 2 \left (
\partial_{m} \Phi \right ) \left (
\partial^{m} \Phi \right )  \right ] \gamma_{ij} \\ & \ \ \ \ + 4 \Phi
\left ( \partial_i \partial_j \Phi \right ) + 2 \left ( \partial_i
\Phi \right ) \left ( \partial_j \Phi \right ),
\end{split}
\end{equation}
which are the quadratic first order scalar perturbations.

The second order trace reversed space-space part of the energy
momentum tensor is
\begin{equation}
\delta^{(2)} \Upsilon_{ij} = a^2 \left [ - \left ( 1 - w \right )
\Phi \delta^{(1)} \rho  \gamma_{ij} + (1 + w) \rho \left (
\partial_{i} \upsilon_{(1)} \right ) \left ( \partial_{j}
\upsilon_{(1)} \right ) \right ].
\end{equation}
It is obvious that we require the following zeroth and first order
equations,
\[
\mathcal{H}^2 = \frac{8}{3} \pi G a^2 \rho,
\]
and
\[
\partial_{i} \upsilon_{(1)} = - \frac{1}{4 \pi G a^2 \rho (1 + w)}
\left [ \partial_i \Phi' + \mathcal{H} \left ( \partial_i \Phi
\right ) \right ].
\]
Note that the terms with $\gamma_{ij}$ as a coefficient will not
play a role since the $\gamma_{ij}$ terms are eliminated by the
operator $\hat{\mathcal{V}}_{i}^{lm}$ defined below. Due to the
limited quantities we keep in our metric the tensorial equations we
calculate are only valid for vector modes, and so these must be
projected out.

We define the Fourier transform of the vector perturbation as
\begin{equation}
S_i(\mathbf{x}, \eta) = \frac{1}{(2 \pi)^{3/2}} \int d^3
\mathbf{k} \left [ S(\mathbf{k}, \eta) e_i(\mathbf{k}) +
\bar{S}(\mathbf{k}, \eta) \bar e_i(\mathbf{k}) \right ]{e}^{i
\mathbf{k}\cdot\mathbf{x}},
\end{equation}
where the two orthonormal basis vectors $\mathbf{e}$ and
$\bar{\mathbf{e}}$ are orthogonal to $\mathbf{k}$. We shall use the
operator $\hat{\mathcal{V}}_{i}^{lm}$ to extract out the
divergenceless vector from a rank-2 tensor
\begin{equation}
\hat{\mathcal{V}}_{i}^{lm} = -\frac{2 i}{(2 \pi)^3} \int d^3
\mathbf{k}'\ {k'^{-2}} \int d^3 \mathbf{x}'\ {k'^{l}} \left [
e_{i}(\mathbf{k}') e^{m}(\mathbf{k}') + \bar{e}_{i}(\mathbf{k}')
\bar{e}^{m}(\mathbf{k}') \right ]  e^{i
\mathbf{k}'\cdot(\mathbf{x} - \mathbf{x}')}.
\end{equation}
This operator will remove any rank-2 tensor which is constructed
from derivatives of a scalar potential. Specifically, any second
order scalar perturbations will be removed and therefore such
perturbations have been neglected in our analysis. The action of
this operator is to produce a rank-1 vector which is a pure vector
mode. Further details concerning the operator can be found
in Appendix~\ref{App1}.

We can then obtain, from the $i-j$ component of the EFE's, the
evolution equation for the second order vector perturbations
\begin{equation}
\hat{\mathcal{V}}_{i}^{lm} \left (\partial_{(l}S_{m)}'+
2\mathcal{H}\partial_{(l}S_{m)}\right ) =
2\hat{\mathcal{V}}_{i}^{lm} \Sigma_{lm},
\end{equation}
where the source term is given by
\begin{equation}
\begin{split}
\Sigma_{lm} & = - 4 \Phi \left ( \partial_{l} \partial_{m} \Phi
\right ) - \frac{2(1 + 3w)}{3(1 + w)} \left ( \partial_{l} \Phi
\right ) \left ( \partial_{m} \Phi \right ) + \frac{4}{3
\mathcal{H}^2 (1 + w)} \left [ \left (
\partial_{l} \Phi' \right ) \left ( \partial_{m} \Phi' \right ) + 2
\mathcal{H} \left ( \partial_{l} \Phi \right ) \left (
\partial_{m} \Phi' \right ) \right ]. \label{EFESV without v}
\end{split}
\end{equation}
For either polarisation, the evolution equation in Fourier space for
the vector mode becomes
\begin{equation}
S' \left ( \mathbf{k}, \eta \right ) + 2 \mathcal{H}S \left (
\mathbf{k}, \eta \right ) = \Sigma (\mathbf{k}, \eta),
\end{equation}
where the source term $\Sigma (\mathbf{k}, \eta)$ is an appropriate
convolution over the quadratic first-order quantities,
\begin{equation}
\begin{split}
\Sigma (\mathbf{k}, \eta) & = - \frac{4 i}{k^2 (2 \pi)^{3/2}} k^{i}
e^{j}(\mathbf{k}) \int d^3 \mathbf{k}' \left ( k'_{i} k'_{j} \right
) \left [ \frac{10 + 6w}{3(1 + w)} \Phi(\mathbf{k}', \eta)
\Phi(\mathbf{k} - \mathbf{k}', \eta) \right.
\\ & \ \ \ \ \left. + \frac{4}{3 (1 + w) \mathcal{H}^2(\eta)}
\Phi'(\mathbf{k}', \eta) \Phi'(\mathbf{k} - \mathbf{k}', \eta) +
\frac{8}{3(1 + w)\mathcal{H}(\eta)} \Phi(\mathbf{k}', \eta)
\Phi'(\mathbf{k} - \mathbf{k}', \eta) \right ]. \label{source for
wave equation in Fourier space SV}
\end{split}
\end{equation}
The general solution for $S\left ( \mathbf{k}, \eta \right )$ can be
written as
\begin{equation}
S \left ( \mathbf{k}, \eta \right ) = \frac{1}{a^2(\eta)}
\int^{\eta}_{\eta_0} d \tilde{\eta}\ a^2(\tilde{\eta}) \Sigma
\left ( \mathbf{k}, \tilde{\eta} \right ). \label{solution for
vector mode}
\end{equation}
We have set the initial conditions for the vector mode to zero at $\eta=\eta_0$.


\subsubsection{Power spectrum}

The power spectrum of the induced vector mode is defined as
\begin{equation}
\left \langle S^{\ast} \left ( \mathbf{k}_1, \eta \right ) S \left
( \mathbf{k}_2, \eta \right ) \right \rangle = \frac{2 \pi^2}{k^3}
\delta^3 \left ( \mathbf{k}_1 - \mathbf{k}_2 \right )
\mathcal{P}_{\mathcal{V}} (k, \eta). \label{power spectrum for
vector mode}
\end{equation}
Substituting the solution \eqref{solution for vector mode} into
\eqref{power spectrum for vector mode} we find
\begin{equation}
\begin{split}
\left \langle S^{\ast} \left ( \mathbf{k}_1, \eta \right ) S \left
( \mathbf{k}_2, \eta \right ) \right \rangle & =
\frac{1}{a^4(\eta)} \int^{\eta}_{\eta_0} d \tilde{\eta}_2
\int^{\eta}_{\eta_0} d \tilde{\eta}_1\ a^2(\tilde{\eta}_1)
a^2(\tilde{\eta}_2) \left \langle \Sigma^{\ast} \left (
\mathbf{k}_1, \tilde{\eta}_1 \right ) \Sigma \left ( \mathbf{k}_2,
\tilde{\eta}_2 \right ) \right \rangle
\\ & = \frac{16}{(2 \pi)^3 a^4(\eta)} \frac{\left [ k_1^{\ m}
e^{n}(\mathbf{k}_1) \right ] \left [ k_2^{\ i} e^{j}(\mathbf{k}_2)
\right ]}{k_1^2 k_2^2 } \int^{\eta}_{\eta_0} d \tilde{\eta}_2
\int^{\eta}_{\eta_0} d \tilde{\eta}_1\ a^2(\tilde{\eta}_1)
a^2(\tilde{\eta}_2) \\ &
\ \ \ \ \times \int d^3 k'_{1} \int d^3 k'_{2}\ \left (k'_{1m}
k'_{1n} \right ) \left ( k'_{2i} k'_{2j} \right ) \Xi \left (
k'_{1}, |\mathbf{k}_1 - \mathbf{k}'_1|, \tilde{\eta}_1 \right )
\Xi \left (k'_{2}, |\mathbf{k}_2 - \mathbf{k}'_2|, \tilde{\eta}_2
\right ) \\ &
\ \ \ \ \times \left \langle \hat{E}^{\ast}(\mathbf{k}'_1)
\hat{E}^{\ast}(\mathbf{k}_1 - \mathbf{k}'_1)
\hat{E}(\mathbf{k}'_2) \hat{E}(\mathbf{k}_2 - \mathbf{k}'_2)
\right \rangle,
\end{split}
\end{equation}
where
\[
\Xi \left ( \mathcal{K}_1, \mathcal{K}_2,{\eta} \right ) =
\frac{10 + 6w}{3(1 + w)} \Phi(\mathcal{K}_1,{\eta})
\Phi(\mathcal{K}_2,{\eta}) + \frac{4}{3(1 +
w)\mathcal{H}^2(\tilde{\eta})} \Phi'(\mathcal{K}_1,{\eta})
\Phi'(\mathcal{K}_2,{\eta}) + \frac{8}{3(1 +
w)\mathcal{H}(\tilde{\eta})} \Phi(\mathcal{K}_1,{\eta})
\Phi'(\mathcal{K}_2,{\eta}).
\]
However, Wick's theorem tells us that
\begin{equation}
\begin{split}
\left \langle \hat{E}^{\ast}(\mathbf{k}'_1)
\hat{E}^{\ast}(\mathbf{k}_1 - \mathbf{k}'_1)
\hat{E}(\mathbf{k}'_2) \hat{E}(\mathbf{k}_2 - \mathbf{k}'_2)
\right \rangle
& = \left \langle \hat{E}^{\ast}(\mathbf{k}'_1)
\hat{E}^{\ast}(\mathbf{k}_1 - \mathbf{k}'_1) \right \rangle \left
\langle \hat{E}(\mathbf{k}'_2) \hat{E}(\mathbf{k}_2 -
\mathbf{k}'_2) \right \rangle \\ &
\ \ \ \ + \left \langle \hat{E}^{\ast}(\mathbf{k}'_1)
\hat{E}(\mathbf{k}'_2) \right \rangle \left \langle
\hat{E}^{\ast}(\mathbf{k}_1 - \mathbf{k}'_1) \hat{E}(\mathbf{k}_2
- \mathbf{k}'_2) \right \rangle \\ &
\ \ \ \ + \left \langle \hat{E}^{\ast}(\mathbf{k}'_1)
\hat{E}(\mathbf{k}_2 - \mathbf{k}'_2) \right \rangle \left \langle
\hat{E}^{\ast}(\mathbf{k}_1 - \mathbf{k}'_1)
\hat{E}(\mathbf{k}'_2) \right \rangle.
\end{split}
\end{equation}
Therefore, the power spectrum of the induced vector mode is
\begin{equation}
\begin{split}
\mathcal{P}_\mathcal{V} \left ( k, \eta \right ) = \frac{1}{k
\pi^5 a^4(\eta)} \int^{\eta}_{\eta_0} d \tilde{\eta}_2 &
\int^{\eta}_{\eta_0} d \tilde{\eta}_1\ a^2(\tilde{\eta}_1)
a^2(\tilde{\eta}_2) \int d^3 k' \left ( k^{a} k'_{a}\right ) \left
[ e^{b}(\mathbf{k}) k'_{b}\right ] \left [ e^{j}(\mathbf{k})
k'_{j}\right ] \\ & \times \Xi \left ( k', |\mathbf{k} -
\mathbf{k}'|, \tilde{\eta}_1 \right ) \left [ \left ( k^{i} k'_{i}
\right ) \Xi \left ( k', |\mathbf{k} - \mathbf{k}'|,
\tilde{\eta}_2 \right ) + \left ( k^{i} k'_{i} - k^2 \right ) \Xi
\left ( |\mathbf{k} - \mathbf{k}'|, k', \tilde{\eta}_2 \right )
\right ]. \label{power spectrum before projection SV general}
\end{split}
\end{equation}
In order to compute the integrals over Fourier space,
we first introduce the dimensionless variables $u$ and $v$, where
\[
v = \frac{k'}{k}\ \mbox{and}\ u = \sqrt{1 + v^2 - 2 v \cos
\theta}.
\]
If we rewrite equation \eqref{power spectrum before projection SV
general} using spherical coordinates in Fourier space, we can carry
out the azimuthal integral trivially. Using the two new variables,
the power spectrum then becomes
\begin{equation}
\begin{split}
\mathcal{P}_\mathcal{V} ( k, \eta ) & = \frac{k^8}{16 \pi^4
a^4(\eta)} \int^{\eta}_{\eta_0} d \tilde{\eta}_2
\int^{\eta}_{\eta_0} d \tilde{\eta}_1\ a^2(\tilde{\eta}_1)
a^2(\tilde{\eta}_2) \int^{\infty}_{0} d v\ \int^{v + 1}_{|v - 1|} d
u\ (uv) \left (v^2 + 1 - u^2 \right ) \left [ (u^2 - 1 - v^2)^2 - 4
v^2 \right ]
\\ & \ \ \ \ \ \times \Xi \left ( kv, ku, \tilde{\eta}_1 \right ) \left \{
( u^2 - 1 - v^2 ) \Xi \left ( kv, ku, \tilde{\eta}_2 \right ) + ( u^2
+ 1 - v^2 ) \Xi \left ( ku, kv, \tilde{\eta}_2 \right ) \right \}.
\end{split}
\end{equation}
The power spectrum can now be calculated once the power spectra
(initial conditions) for the scalar modes are chosen.

\section{VECTOR MODE POWER SPECTRA}

We shall now investigate the power spectrum of the induced vector modes
during the radiation era.

After substituting for the first order solution for $\Phi$ for the
radiation era, the power spectrum then becomes
\begin{equation}\label{PS_rad}
\begin{split}
\mathcal{P}_{\mathcal{V}} (k, \eta) & = \frac{(243)^2}{4 (k
\eta)^{4}} \int^{\infty}_{0} d v \int^{v + 1}_{|v - 1|} d u\
\mathcal{P}_{\Phi}(k u) \mathcal{P}_{\Phi}(k v) {\cal F}(u,v,x),
\end{split}
\end{equation}
where
\begin{equation}
\begin{split}
{\cal F}(u,v,x) & = \frac{1}{(u v)^8} \left ( v^2 + 1 - u^2 \right )
\left [ (u^2 - 1 - v^2)^2 - 4 v^2 \right ]
\\ & \ \ \ \ \times \int^{x}_{x_0} d \tilde{x}_{1}\ \mathcal{I}_1 (\tilde{x}_{1})
\left [ (u^2 - 1 - v^2) \int^{x}_{x_0} d \tilde{x}_{2}\
\mathcal{I}_1 (\tilde{x}_{2}) + (u^2 + 1 - v^2) \int^{x}_{x_0} d
\tilde{x}_{2}\ \mathcal{I}_2 (\tilde{x}_{2}) \right ],\label{power
spectrum in radiation}
\end{split}
\end{equation}
and $x$ is another dimensionless variable defined as $x = k \eta$, and $x_0=k\eta_0$.
We have defined the functions
\begin{equation}
\mathcal{I}_j (x) = \sum^{5}_{m = 1} \sum^{4}_{n = 1}\ \sin \left
( \alpha_n x + \phi_n \right ) \frac{M^{j}_{nm}}{x^{m - 1}},
\end{equation}
with the coefficients $\alpha_n$, $\phi_n$ and $M^{j}_{nm}$ defined as
\begin{equation}
\phi_{n} = \frac{\pi}{2} \left ( \begin{array}{c}
\mathbf{1} \\
\mathbf{0}
\end{array} \right )\
\mbox{and}\ \alpha_n =\frac{1}{\sqrt{3}} \left ( \begin{array}{c}
-u \mathbf{1} + \upsilon \mathbf{b} \\
u \mathbf{1} + \upsilon \mathbf{a}
\end{array} \right ),
\end{equation}
\begin{equation}
M^{1}_{nm} = \left ( \begin{array}{ccccc} \frac{u^2
\upsilon^2}{18}\mathbf{b} & \mathbf{0} & (\frac{1}{6} u^2 +
\frac{1}{2} \upsilon^2) \mathbf{a} + u \upsilon \mathbf{1} &
\mathbf{0} & 3 \mathbf{b} \\
\mathbf{0} & \frac{u^2 \upsilon}{6 \sqrt{3}} \mathbf{1} + \frac{u
\upsilon^2}{2 \sqrt{3}} \mathbf{a} & \mathbf{0} & - \sqrt{3}
\upsilon \mathbf{1} + \sqrt{3} u \mathbf{b} & \mathbf{0}
\end{array} \right ),
\end{equation}
\begin{equation}
M^{2}_{nm} = \left ( \begin{array}{ccccc} \frac{u^2
\upsilon^2}{18}\mathbf{b} & \mathbf{0} & (\frac{1}{2} u^2 +
\frac{1}{6} \upsilon^2) \mathbf{a} + u \upsilon \mathbf{1} &
\mathbf{0} & 3 \mathbf{b} \\
\mathbf{0} & \frac{u^2 \upsilon}{2 \sqrt{3}} \mathbf{1} + \frac{u
\upsilon^2}{6 \sqrt{3}} \mathbf{a} & \mathbf{0} & - \sqrt{3}
\upsilon \mathbf{1} + \sqrt{3} u \mathbf{b} & \mathbf{0}
\end{array} \right ).
\end{equation}
Here we have defined the matricies $\mathbf{1}$, $\mathbf{0}$,
$\mathbf{a}$ and $\mathbf{b}$ as
\begin{equation}
\mathbf{1} = \left ( \begin{array}{c}
1 \\
1
\end{array} \right ),\
\mathbf{0} = \left ( \begin{array}{c}
0 \\
0
\end{array} \right ),\
\mathbf{a} = \left ( \begin{array}{c}
1 \\
-1
\end{array} \right ),\
\mathbf{b} = \left ( \begin{array}{c}
-1 \\
1
\end{array} \right ).
\end{equation}
As we have four integrals to carry out it is useful to calculate
them analytically where possible. We can do this for the
$x$-integrals to get
\begin{equation}
\begin{split}
\int^{x}_{x_0} d \tilde{x}\ \mathcal{I}_j (\tilde{x}) = &
\sum^{5}_{m = 1} \sum^{4}_{n = 1}\ M^{j}_{nm} \left \{ \left [
\sum^{m - 3}_{k = 1} \frac{(m - k - 3)!}{(m - 2)!} \alpha^{\ k}_{n}
\sin \left( \alpha_{n} \tilde{x} + \phi_n + \frac{k + 2}{2} \pi
\right ) \tilde{x}^{(2 + k - m)} \right ]^{x}_{x_0} \right. \\
& \left. - \frac{\alpha^{\ (m - 2)}_{n}}{(m - 2)!} \left [
\mbox{Si} \left(\alpha_n \tilde{x} \right) \cos \left(\phi_n +
\frac{m}{2} \pi \right) + \mbox{Ci} \left(\alpha_n \tilde{x}
\right) \sin \left(\phi_n + \frac{m}{2} \pi \right) \right
]^{x}_{x_0} \right \}.
\end{split}
\end{equation}
For the radiation era we assume that all modes are well outside the
horizon when the interaction begins and therefore can set $x_{0} =0$.

\subsection{Interaction of scalar modes}

Before calculating the power spectrum for the case of power law
scalar modes, it is useful to investigate how the vector modes are
generated from individual scalar modes. It has been shown that a
single scalar mode with an isotropic distribution will induce second-order gravitational
waves~\cite{ACW}. 
This is not the case with vector modes: 
scalar modes of differing wavelengths need to interact to generate
vector modes, as we shall see. To investigate this we choose then a scalar power
spectrum of the form
\begin{equation}
\mathcal{P}_{\Phi}(k) =\frac{4}{9} \mathcal{A}^2
\Delta_{\mathcal{R}}^2(k_{CMB})
\left\{\delta[\ln(k_{1}/k)]+\delta[\ln(k_{2}/k)]\right\},
\end{equation}
where $\mathcal{A}$ is the mean amplitude of each wavenumber,
$k_{i}$, relative to the observed amplitude of the primordial power
spectrum, $\Delta^2_{\mathcal{R}}(k_{CMB})$, at wavenumber
$k_{CMB}\gg k_{i}$. We assume for simplicity that they both have the
same amplitude. Carrying out the $u$ and $v$ integrals, we then find that the vector power spectra becomes,
in terms of $v_i=k_i/k$,
\begin{equation}
\mathcal{P}_{\mathcal{V}} (k, \eta) = 26244 \mathcal{A}^4
\Delta_{\mathcal{R}}^4 \frac{k_1k_2}{x^4k^2}
\left[\mathcal{F}(v_1,v_2,x) + \mathcal{F}(v_2,v_1,x)\right]
\end{equation}
provided $v_1+1>v_2>|v_1-1|$, and is zero otherwise. Therefore modes
are induced for all wavenumbers $k$ such that $k_1+k_2>k>|k_1-k_2|$, and are scattered into angles such that $u_1=v_2$, and $u_2=v_1$ (a further requirement from  carrying out the integration). In the case where only one input mode is present, so $v_2=v_1$, this inequality becomes $k_1>k/2$ while we also have $v_1=u_1$, as in \cite{ACW}; from Eq.~(\ref{power spectrum in radiation}) with $u=v$, however, we see that $\mathcal{F}$ vanishes in this case. Therefore, we can see that vector modes cannot be induced by a single scalar degree of freedom. The physical reason for this is because vector modes are associated with rotational degrees of freedom. A consequence of $k_1=k_2$ is that $ \theta_1=\pm\theta_2$, i.e., the input modes only have momentum along the same axis in Fourier space. Consequently,  there is no angular momentum generated, and hence no vectors.

Provided that $k_1\neq k_2$, we can have vectors induced over the appropriate range of wavelengths. Closely separated scalar modes will produce a much broader spectrum
of vector modes while modes of vastly differing wavelengths will
produce a very narrow range of vectors, with wavenumbers close to
the largest input wavenumber. Note that as the generated wavenumbers are
restricted from above and below, we can't expect any noise on large
scales, as is the case for gravitational waves. This is also evidenced
by the fact that one input mode can't produce any vectors~-- there would be nothing to set the long wavelength cutoff in that case.

\begin{figure}[th!]
\begin{center}
\includegraphics[width=0.8\textwidth]{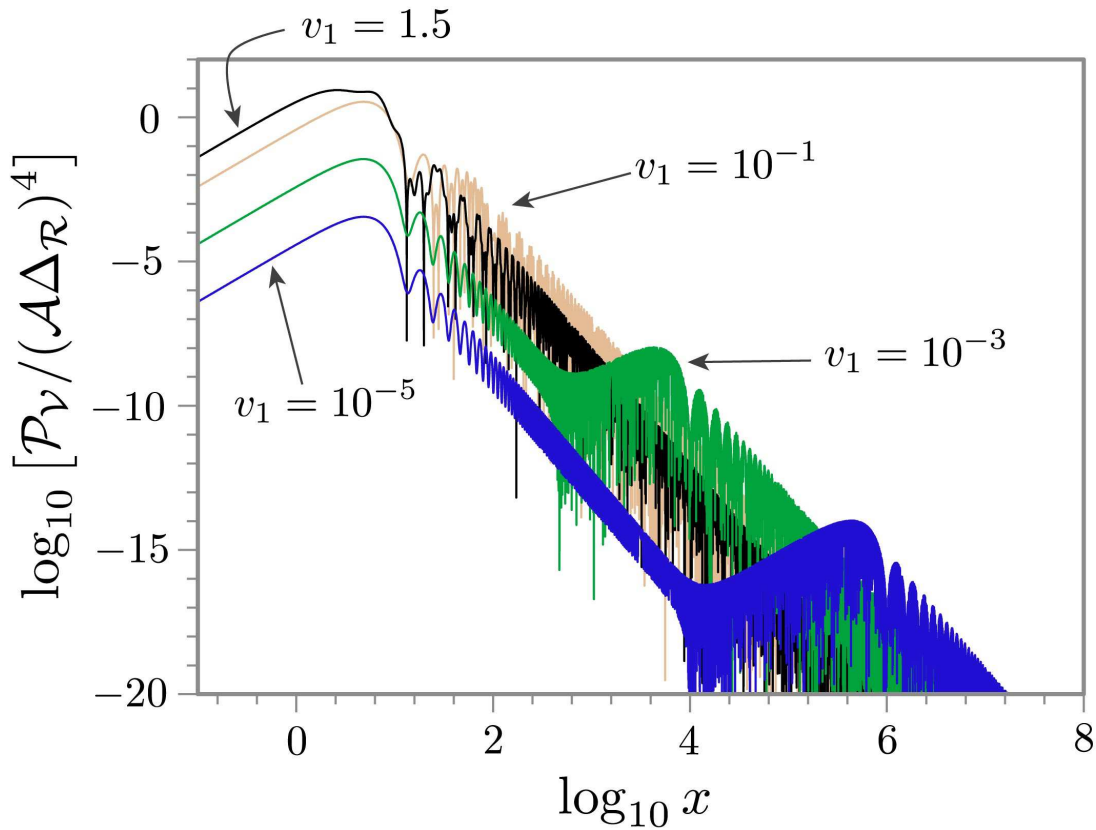}
\caption{The power spectrum of vector modes induced by two
interacting scalar modes. Although maximum power is generated in the scenario
$k_1/k_2\sim1$ shortly after the Hubble radius is crossed, at late times
scalar interactions with vastly differing wavelengths produce more
power once the long-wavelength mode enters the Hubble-radius. }
\label{fig-df1}
\end{center}
\end{figure}

In Fig.~\ref{fig-df1} we show the induced vector modes as a function
of $x$ for various $v_1$ with $v_2=1$ (so we require $0<v_1<2$),
i.e., the evolution of modes of wavenumber $k=k_2$. 
While the generated mode is outside the Hubble radius, there is
power-law growth, with $\mathcal{P_V}\sim \eta^2$. When $k_2$ enters
the Hubble radius, the principle generation of vector modes stops
shortly thereafter, and the induced modes start to decay as
$\eta^{-4}$. This continues until the longer wavelength mode enters
the Hubble radius at $k_1\eta=1\Rightarrow x\sim1/v_1$. This then
generates a further burst of vector modes, which we can see by
progressively more pronounced knees, as $v_1\to0$, in the curves at
late times. For the case when $v_1=1.5$, on the other hand, we see
some confusion as the modes enter the horizon more-or-less together,
before decaying as normal. 

Thus we see that vector modes are only efficiently generated when at least one of the scalar modes is entering its Hubble radius~-- provided another scalar mode exists to help seed the vector mode. This explains why we have knees in the evolution of the generated vector modes, since two interacting scalars enter the Hubble radius at different times.  The power generated into vectors  as each mode enters depends on the relative ratio $k_1/k_2$. Modes of similar wavelength generate more overall power, because they are entering at the same time; modes that are widely separated in wavenumber don't generate as much overall power but produce more pronounced knees instead.

\subsection{Power law scalar modes}

Let us now investigate the spectrum of vector modes from power-law
scalar modes. To do this, we assume that the input power spectrum
is:
\begin{eqnarray}
\mathcal{P}_{\Phi}(k) = \frac{4}{9}
\Delta_{\mathcal{R}}^2~\left(\frac{ k}{k_{CMB}}\right)^{n_{s}-1},
\label{power law PS}
\end{eqnarray}
where the index $n_s$ tells us the tilt of the spectrum relative to
scale-invariance, $n_s=1$, and $k_{CMB}$ is a pivot scale for the
power spectrum~\cite{WMAP3}. The induced vector modes are then given
by
\begin{equation}
\mathcal{P}_{\mathcal{V}} (k, \eta) = \frac{729
\Delta_{\mathcal{R}}^4}{(k \eta)^4} \left(\frac{
k}{k_{CMB}}\right)^{2(n_{s}-1)} \mathcal{F}_{n_{s}}(x),
\end{equation}
where $\mathcal{F}_{n_{s}}(x)$ is defined as
\begin{equation}
\mathcal{F}_{n_{s}}(x) = \int^{\infty}_{0}\ d v\ \int^{v + 1}_{|v
- 1|}\ d u\ (u v)^{n_{s}-1} \mathcal{F}(u,v,x).
\end{equation}

\begin{figure}[th!]
\begin{center}
\includegraphics[width=0.6\textwidth]{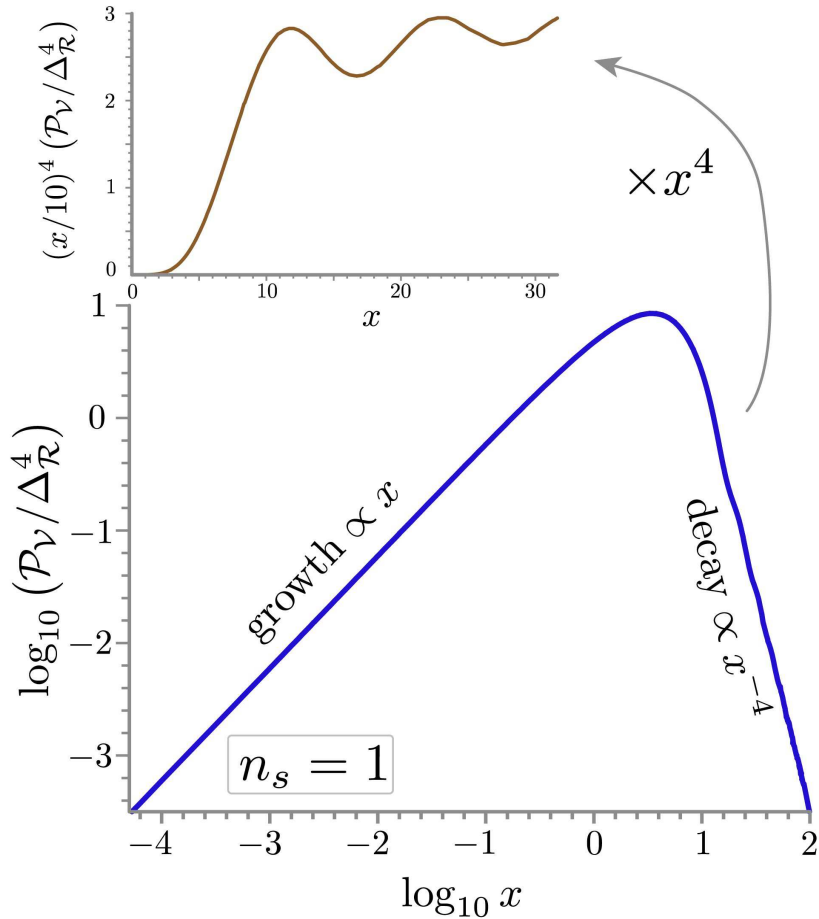}
\caption{The power spectrum of vector modes induced by
scale-invariant scalar modes. Scalar modes outside the Hubble radius
interact to give power-law growth at until the modes enter the
Hubble radius. Thereafter the modes decay as normal vectors, with
some gentle oscillatory features.} \label{fig-pl-rad}
\end{center}
\end{figure}

We integrate this numerically, and show the results in
Fig.~\ref{fig-pl-rad} for the case $n_s=1$.
The tilt of the scalar power spectrum tends to affect the amplitude
of second-order modes on large scales at the level of a few
percent~\cite{ACW, Mol1}. Viewing $x$ as time for constant $k$, we
see that the modes grow as $\eta$, peak when inside the Hubble
radius and decay as $\eta^{-4}$. While the modes are decaying there
are faint oscillations as shown in the top panel of the Figure. This
Figure can also be interpreted as the power spectrum at fixed time,
showing the usual features.

It is worth mentioning that we have taken the upper limit of the
$k'$ integral to be infinity. In reality there is a cutoff from
the end of inflation, at $\eta=\eta_*$, corresponding to modes
which are inside the Hubble radius at that time, $k_*=1/\eta_*$,
so giving a finite upper limit to the $v$-integral, $v_*= k_*/k$ (although this is on very small scales in reality).
This causes a break from linear scaling in $x$ in the power spectrum for $x\lesssim 1/v_*$,
and we may analytically find the leading behavior of Eq.~(\ref{PS_rad}) is $\mathcal{P_V}\sim \frac{32}{15}v_* x^2$. Why is this the case?

In the interacting delta function case we saw that modes are
efficiently produced~-- and grow like $x^2$~-- when both modes are
outside the Hubble radius; once one is inside and the other outside
there is effectively no generation of vectors. In power-law case,
then, the modes which are generating vectors are those outside the
Hubble radius, providing an effective cutoff to the $v$-integral of
$v\sim 1/x$, so giving us growth $\propto x$, when $v_*\gg 1/x$.
When we have the cutoff $v_*$ on the other hand, for early times
when $\eta<1/k_*$, all relevant modes are outside the Hubble radius,
and interact coherently giving us growth $\propto \eta^2$. For
$\eta>1/k_*$, modes which have entered the Hubble radius no longer
contribute to the generation of modes outside the Hubble radius
giving weaker growth $\propto x$. Of course, we are not in a
position here to analyse times before inflation ends, but this helps
us understand why we have the $x$-scaling behaviour we do.

\section{CONCLUSIONS}

An important feature of inflation is the lack of vector modes: if
they were observed to have a similar spectrum and amplitude to the
scalars then this could prove difficult for inflation, and lend
favor to other theories of the early universe, e.g. Pre Big Bang
scenarios and Ekpyrotic models~\cite{Bat, Mena,Bojowald}. However,
there is a $\chi^2$-distribution of vectors produced by inflation
as a consequence of the non-linear interaction of scalar modes,
which has received relatively little attention to date.

We have investigated the generation of vector modes induced by
primordial density perturbations during the radiation dominated
era. Performing a perturbative expansion to second order, we
isolated the scalar terms which source the vector perturbations.
We then calculated the power spectrum of the metric vector mode,
and analysed its form. In order to understand the generation of
modes we investigated individual scalar modes generating vectors,
and demonstrated that, contrary to the case of gravitational
waves, vector modes can't be generated by an isotropic
distribution of scalars of a single wavelength, owing to the
spin-1 nature of vector modes: rotational degrees of freedom must be generated by scattering of non-parallel input modes. We then demonstrated that  vectors
are generated by modes of differing wavelength whenever one of the
two scalar modes is entering the Hubble radius. The amplitude of
the generated modes depends on the ratio of input wavenumbers;
maximum power is generated when the modes are not too widely
separated. After investigating the generation of modes, we then
presented the power spectrum for scale-invariant scalar modes,
displaying our results in terms of the variable $x=k\eta$: i.e.,
they may be interpreted the temporal evolution of a single scalar
mode, or the power at a fixed time. Interestingly the maximum
power generated is the same at all times, but the position of this
peak changes with wavelength, such that~$x\sim1 \Rightarrow
k\sim1/\eta$. This is due to the fact that the modes are
efficiently generated as they enter the Hubble radius, and aren't
generated significantly while outside.

There are some open questions raised by the study presented here. 
In particular, it is not clear how the power spectrum for $S_i$ we 
have calculated will be related to observable quantities. It is 
gauge invariant so must be observable, by the results presented 
in~\cite{BS}; it also represents all possible degrees of freedom 
of vectors generated by scalars, under the conditions laid out 
here. Thus, although there may be a `better' variable, it must be 
related to $S_i$ by quadrature (plus some further scalar-squared 
contributions). The effects we have presented here will have interesting
implications for a variety of phenomena such as the CMB; the issue
of how significant is left for future work.

\acknowledgements

We would like to thank Bruce Bassett, Rob Crittenden, Ruth Durrer, Roy Maartens, Jean-Phillipe Uzan and David Wands for useful discussions and comments. THCL and CC acknowledge
financial support from the University of Cape Town; all authors are
supported by the National Research Foundation (South Africa). KK is
additionally supported by the Italian {\it Ministero Degli Affari
Esteri-DG per la Promozione e Cooperazione Culturale} under the
joint Italy/ South Africa Science and Technology agreement.

\appendix

\begin{widetext}
\section{The extraction operator}\label{App1}

In this section we consider the extraction operator discussed in
this paper. We start by defining the Fourier basis used for the
purposes of harmonic decompositions. An arbitrary scalar in real
space can be expressed as a Fourier integral
\be
S(\mathbf{x},\eta) = \frac{1}{(2\pi)^{3/2}} \int d^3
\mathbf{k}~S(\mathbf{k},\eta) e^{i\mathbf{k}\cdot\mathbf{x}} .
\ee
A divergence free vector in real space can then be expressed as a
Fourier integral
\be
V_{a}(\mathbf{x},\eta) = \frac{1}{(2\pi)^{3/2}} \int d^3
\mathbf{k}~ \left[ \bar{V}(\mathbf{k},\eta)\bar{e}_{a}(\mathbf{k}) +
{V}(\mathbf{k},\eta){e}_{a}(\mathbf{k}) \right]
e^{i\mathbf{k}\cdot\mathbf{x}} ,
\ee
where $e_{a}(\mathbf{k})$ and $\bar{e}_{a}(\mathbf{k})$ are orthogonal parity vectors, which are also orthogonal to $\mathbf{k}$. 
Similarly, a transverse traceless tensor (a tensor mode) in real space can be expressed as
\be
T_{ab}(\mathbf{x},\eta) = \frac{1}{(2\pi)^{3/2}} \int d^3
\mathbf{k}~\left[ {T}(\mathbf{k},\eta){q}_{ab}(\mathbf{k}) +
\bar{T}(\mathbf{k},\eta)\bar{q}_{ab}(\mathbf{k}) \right]
e^{i\mathbf{k}\cdot\mathbf{x}},
\ee
where the two polarization tensors $q_{ab}$ and $\bar{q}_{ab}$ can
also be expressed in terms of the parity vectors $e_{a}$ and $\bar{e}_{a}$, and are orthogonal to $\mathbf{k}$.
An arbitrary symmetric trace-free spatial tensor in real space
\be\label{RealIn}
A_{ab}(\mathbf{x},\eta) = \left[ \partial_{a}\partial_{b}
-\frac{1}{3}\gamma_{ab}\partial^{c}\partial_{c} \right] A^{(S)}(\mathbf{x},\eta)
+\partial_{(a}A^{(V)}_{b)}(\mathbf{x},\eta)
+A^{(T)}_{ab}(\mathbf{x},\eta),\nonumber\\
\ee

\noindent which has explicit scalar, vector and tensor
contributions. We can also express $A_{ab}$ as a Fourier integral
\begin{eqnarray}\label{FourierIn}
A_{ab}(\mathbf{x},\eta) &=& \frac{1}{(2\pi)^{3/2}} \int d^3
\mathbf{k}~\left\{ -\left[{k}_{a}{k}_{b}
-\frac{1}{3}\gamma_{ab}{k}^2 \right] A^{(S)}(\mathbf{k},\eta)
+{i}A^{(V)}(\mathbf{k},\eta){k}_{(a}{e}_{b)}
\right.\nonumber\\
\nonumber\\
&&~~~~~\left. \frac{}{}+{i}\bar{A}^{(V)}(\mathbf{k},\eta)
{k}_{(a}\bar{e}_{b)} +A^{(T)}(\mathbf{k},\eta){q}_{ab}
+\bar{A}^{(T)}(\mathbf{k},\eta)\bar{q}_{ab} \right\}
e^{i\mathbf{k}\cdot\mathbf{x}} .
\end{eqnarray}
The even-parity vector contribution can then be extracted in Fourier space by applying the following
operator
\begin{equation}
\hat{\mathcal{V}}_{i}^{lm} = -\frac{2 i}{(2 \pi)^3} \int d^3
\mathbf{k}'\ {k'^{-2}} \int d^3 \mathbf{x}'\ {k'^{l}} \left [
e_{i}(\mathbf{k}') e^{m}(\mathbf{k}') + \bar{e}_{i}(\mathbf{k}')
\bar{e}^{m}(\mathbf{k}') \right ]  e^{i \mathbf{k}'\cdot(\mathbf{x}
- \mathbf{x}')}.
\end{equation}
\noindent Applying the operator $\hat{\mathcal{V}}_{l}{}^{ab}$ to
Eq.~\ref{RealIn} gives
\begin{eqnarray}
A^{(V)}_{l}(\mathbf{x},\eta) &=&\hat{\mathcal{V}}_{l}{}^{ab}A_{ab}\nonumber\\
\nonumber\\
&=&-\frac{2 i}{(2 \pi)^3} \int d^3 \mathbf{k}'\ {k'^{-2}} \int d^3
\mathbf{x}'\ {k'^{a}} \left [ e_{l}(\mathbf{k}') e^{b}(\mathbf{k}')
+ \bar{e}_{l}(\mathbf{k}') \bar{e}^{b}(\mathbf{k}') \right ]  e^{i
\mathbf{k}'\cdot(\mathbf{x} - \mathbf{x}')} A_{ab}(\mathbf{x}',\eta)
.
\end{eqnarray}
The extraction of the vector component is made possible by taking
advantage of the various properties of both the parity vectors
($e_{a}$ and $\bar{e}_{a}$) and the polarization tensors ($q_{ab}$
and $\bar{q}_{ab}$). We now consider as an example
one possible contribution from the first order squared terms. For
simplicity we start with contributions from a term made up of the product of first
order scalars and do not reconstruct in real space, looking
only at the Fourier amplitudes. Consider a term of the type
\be
\Phi\partial_{a}\partial_{b}\Phi = \frac{1}{(2\pi)^{3}} \left[ \int d^3
\mathbf{k}_{1}~\Phi(\mathbf{k}_{1},\eta)
e^{i\mathbf{k}_{1}\cdot\mathbf{x}} \right]\left[ \int d^3
\mathbf{k}_{2}~\Phi(\mathbf{k}_{2},\eta)
\left(-{k}_{2a}{k}_{2b} \right)
e^{i\mathbf{k}_{2}\cdot\mathbf{x}} \right].
\ee
The Fourier amplitude of the vector part of this is then
\be
[\Phi\partial_{a}\partial_{b}\Phi]^{(V)}(\mathbf{k})=\frac{2i}{(2\pi)^{3/2}} \int d^3
\mathbf{k}_{2}~ \Phi(\mathbf{k}-\mathbf{k}_{2},\eta)
~\Phi(\mathbf{k}_{2},\eta)~\frac{{k}^a \left[{e}^b\left(\mathbf{k}\right) +
\bar{e}^b\left(\mathbf{k}\right) \right]
{k}_{2a}{k}_{2b}}{{k}^2},
\ee
where we have carried out a
real space integral and a $k$-space integral.
\end{widetext}



\end{document}